\documentclass{iaus}
\usepackage{graphics,epsfig}

  \checkfont{eurm10}
  \iffontfound
    \IfFileExists{upmath.sty}
      {\typeout{^^JFound AMS Euler Roman fonts on the system,
                   using the 'upmath' package.^^J}%
       \usepackage{upmath}}
      {\typeout{^^JFound AMS Euler Roman fonts on the system, but you
                   dont seem to have the}%
       \typeout{'upmath' package installed. iaus.cls can take advantage
                 of these fonts,^^Jif you use 'upmath' package.^^J}%
      }
  \else
  \fi

  \checkfont{msam10}
  \iffontfound
    \IfFileExists{amssymb.sty}
      {\typeout{^^JFound AMS Symbol fonts on the system, using the
                'amssymb' package.^^J}%
       \usepackage{amssymb}%

      }{}
  \fi

  \IfFileExists{amsbsy.sty}
    {\typeout{^^JFound the 'amsbsy' package on the system, using it.^^J}%
     \usepackage{amsbsy}}
    {\providecommand\boldsymbol[1]{\mbox{\boldmath $##1$}}}

\def\REdd{\dot M/\dot M_{\rm Edd}}
\def\lsim{\ifmmode{\lower0.3em\hbox{$\,\buildrel <\over\sim\,$}}\else{\lower0.3em\hbox{\,\buildrel <\over\sim\,}}\fi}
\def\gsim{\ifmmode{\lower0.3em\hbox{$\,\buildrel >\over\sim\,$}}\else{\lower0.3em\hbox{\,\buildrel >\over\sim\,}}\fi}
\def\kms{\ifmmode{~{\rm km~s^{-1}}}\else{~km~s$^{-1}$}\fi}
\def\Msol{\ifmmode{~{\rm M_{\odot}}}\else{M$_{\odot}$}\fi}
\def\ion#1#2{\hbox{#1{\sc\,#2}}}

\newsavebox{\astrutbox}
\sbox{\astrutbox}{\rule[-5pt]{0pt}{20pt}}

\title[UV Spectra of Double-Peaked Emitters]{The Ultraviolet Spectra of Active Galaxies With Double-Peaked
Emission Lines}

\author[M. Eracleous {\it et al.\/}]%
{M. Eracleous$^1$
J. P. Halpern$^2$, 
T. Storchi-Bergmann$^3$, 
A. V. Filippenko$^4$,
A. S. Wilson$^5$,
\& M. Livio$^6$.
}

\affiliation{
$^1$Deptartment of Astronomy \& Astrophysics, The Pennsylvania State 
University, 525 Davey Lab, University Park, PA 16802 \\[\affilskip]
$^2$Department of Astronomy, Columbia University, 550 West 120th Street, 
New York, NY 10027  \\[\affilskip]
$^3$Instituto de Fisica, UFRGS, Campus do Vale, Porto Alegre, 
RS, Brazil \\[\affilskip]
$^4$Dept. of Astronomy, University of California, Berkeley,
CA 94720 \\[\affilskip]
$^5$Dept. of Astronomy, University of Maryland, College Park,
MD 20742 \\[\affilskip]
$^6$Space Telescope Science Institute, 3700 San Martin Drive,
Baltimore, MD 21218
}

\pubyear{2004}
\volume{222}
\pagerange{1--8}
\date{?? and in revised form ??}
\jname{The Interplay among Black Holes, Stars and ISM \\in Galactic Nuclei}
\editors{Th. Storchi Bergmann, L.C. Ho \& H.R. Schmitt, eds.}
\begin{document}

\maketitle

\begin{abstract}
We present the results of UV spectroscopy of AGNs with double-peaked
Balmer emission lines. In 2/3 of the objects, the far-UV resonance
lines are strong, with single-peaked profiles resembling those of
Seyfert galaxies. The \ion{Mg}{ii} line is the only UV line with a
double-peaked profile. In the remaining objects, the far-UV resonance
lines are relatively weak but still single-peaked. The latter group
also displays prominent UV absorption lines, indicative of a
low-ionization absorber. We interpret the difference in the profiles
of the emission lines as resulting from two different regions: a
dense, low-ionization accretion disk (the predominant source of the
Balmer and \ion{Mg}{ii} lines), and a lower density, higher-ionization
wind (the predominant source of the far-UV resonance lines). These
results suggest a way of connecting the double-peaked emitters with
the greater AGN population: in double-peaked emitters the accretion
rate onto the black hole is low, making the wind feeble and allowing
the lines from the underlying disk to shine through. This scenario
also implies that in the majority of AGNs, the wind is the source of
the broad emission lines.
\end{abstract}

\firstsection 

\section{Introduction}

Broad, double-peaked Balmer lines are found in the spectra of
approximately 20--25\% of radio-loud AGNs at $z<0.4$
(\cite{eracleous94}, \cite[2003]{eracleous03a}) and in the spectra of
4\% of {\it all} AGNs at $z<0.33$ from the Sloan Digital Sky Survey
(\cite{strateva03}). Half of these profiles are well described by
kinematic models of axisymmetric, relativistic, Keplerian disks
(\cite{chen89a}; \cite{chen89b}), while the other half require a
modified version of this model in which the disk is not axisymmetric
(\cite{eracleous95}; \cite{strateva03}).  The hosts of double-peaked
emission lines (hereafter the ``double-peaked emitters'') share a
number of distinguishing characteristics which make them stand out
from the average AGN and suggest that these objects represent the
extreme end of the distribution of AGN properties, namely: (a) Balmer
line widths that are, on average, twice as large as those of other
AGNs, (b) steep broad-line Balmer decrements, (c) unusually strong
low-ionization emission lines (i.e., [\ion{O}{i}] and [\ion{S}{ii}]),
as well as unusually large [\ion{O}{i}]/[\ion{O}{iii}] ratios, and (d)
a great deal of starlight in the optical continuum. These
characteristics support the scenario of \cite[Chen \& Halpern
(1989)]{chen89b} in which the inner accretion in these objects is an
ion torus (\cite{rees82}; known today as a radiatively inefficient
accretion flow). Some of the spectroscopic properties of double-peaked
emitters are reminiscent of LINERs and the relation between the two
groups is underscored by the detection of double-peaked Balmer lines
in several LINERs.

\begin{figure}
\centerline{\psfig{file=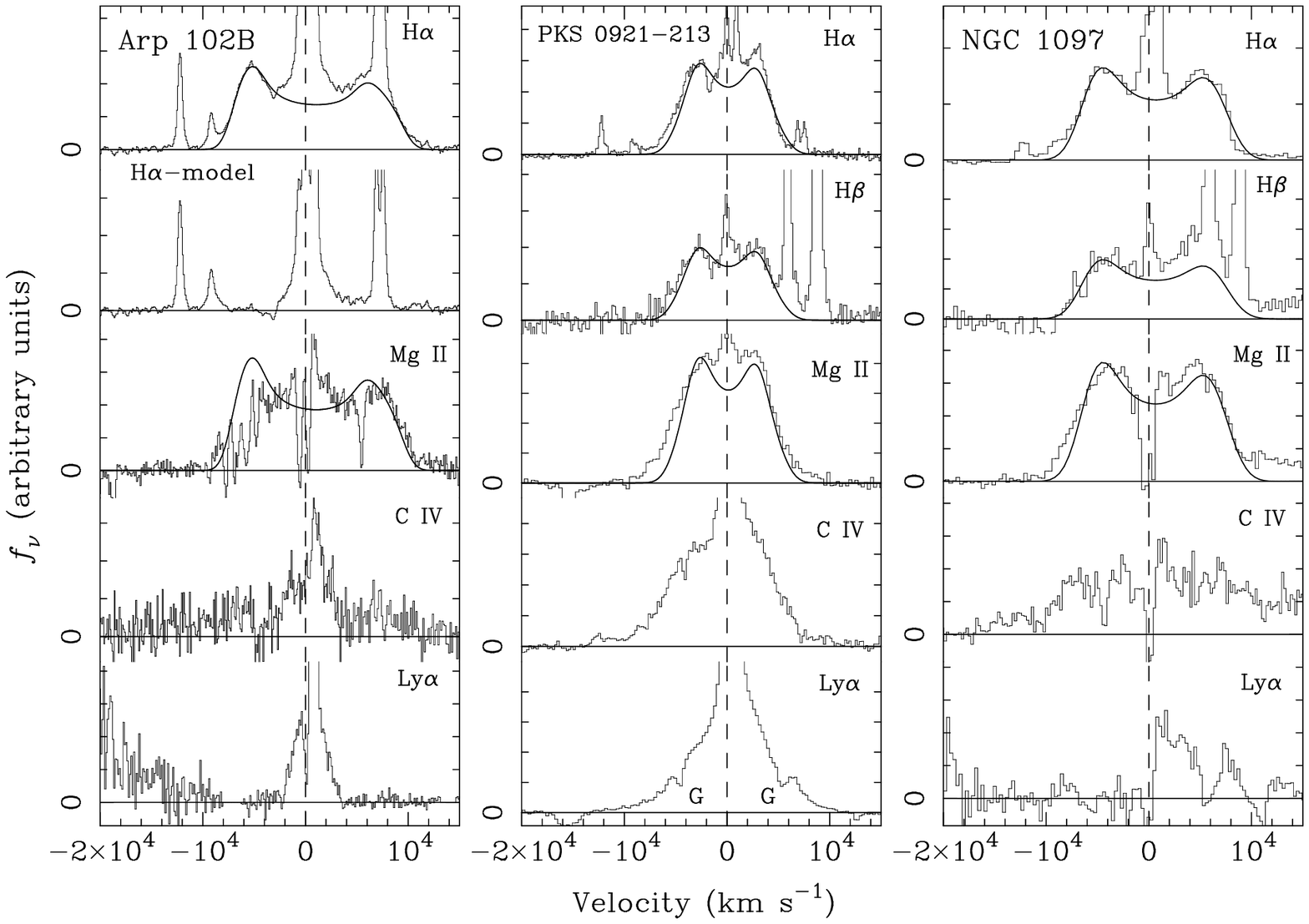,width=4.8in}}
\caption{Optical, near-UV, and far-UV emission-line profiles of three
  double-peaked emitters. A circular-disk model is superposed on the
  Balmer and \ion{Mg}{ii} profiles for comparison. The H$\alpha$
  profile of Arp~102B after subtraction of the model is included, to
  show that the residual resembles the Ly$\alpha$ profile. Absorption
  lines marked with a ``G'' come from the gas in the Milky way, while
  all others arise in gas associated with the AGN.}
\end{figure}

\section{UV Spectroscopy with the Hubble Space Telescope and Interpretation}

UV spectra of several bright double-peaked emitters were observed with
the {\it HST's} FOS and STIS instruments in order to compare the
profiles of the high- and low-ionization UV emission lines with those
of the Balmer lines, as well as to detect and study the shape of the
UV continuum. The results of the observations of the prototypical
object, Arp~102B, were presented and discussed by \cite[Halpern et
al. (1996)]{halpern96}, while work on the other objects is in
progress. A study of the spectral energy distribution of one of these
objects, NGC~1097, is presented by Nemmen et al., in this volume.

The \ion{Mg}{ii} line is the only UV line with a double-peaked
profile, while all the other UV lines are single-peaked.  This is
shown in Figure~1, where we compare the Balmer and UV line profiles of
three representative objects.  Particularly striking is the dramatic
difference between the profiles of the Ly$\alpha$ and Balmer lines.
In 2/3 of the objects (exemplified by PKS~0921--213 in Fig.~1), the
far-UV resonance lines are strong, with single-peaked profiles
resembling those of Seyfert galaxies.  In the remaining objects (e.g.,
Arp~102B and NGC~1097 in Fig.~1), the far-UV resonance lines are
relatively weak but still single-peaked. The emission-line profiles of
Arp~102B also serve to illustrate that there are at least two
broad-emission line regions in the same object. In particular, the
profile of the H$\alpha$ line can be decomposed into a very broad,
double-peaked part which has no counterpart in the Ly$\alpha$ line,
and a bell-shaped part of intermediate width, which closely resembles
the Ly$\alpha$ profile. The double-peaked part can be attributed to
emission from the dense, low-ionization gas in the accretion disk.
The extremely weak Ly$\alpha$ emission from the disk was actually
predicted by the models of weakly photoionized accretion disks by
\cite[Collin-Souffrin \& Dumont (1989) and Dumont \& Collin-Souffrin
(1990)]{collin89,dumont90}: at the high densities and column densities
of the accretion disk Ly$\alpha$ photons are trapped by resonance
scattering and the $n=2$ level of hydrogen is deexcited by collisions,
thus the Ly$\alpha$ photons are effectively lost.  On the other hand
Balmer-line photons can escape relatively easily.  The bell-shaped
component can be ascribed to a medium of lower density and higher
ionization, presumably a wind overlaying the disk.  The propagation of
the line photons through the accelerating wind results in a
non-axisymmetric emissivity pattern as seen by the observer. Thus, the
emission-line profiles from the wind are single-peaked, even though
the dominant motion of the gas is rotational (\cite{murray97}).

\begin{figure}
\centerline{\psfig{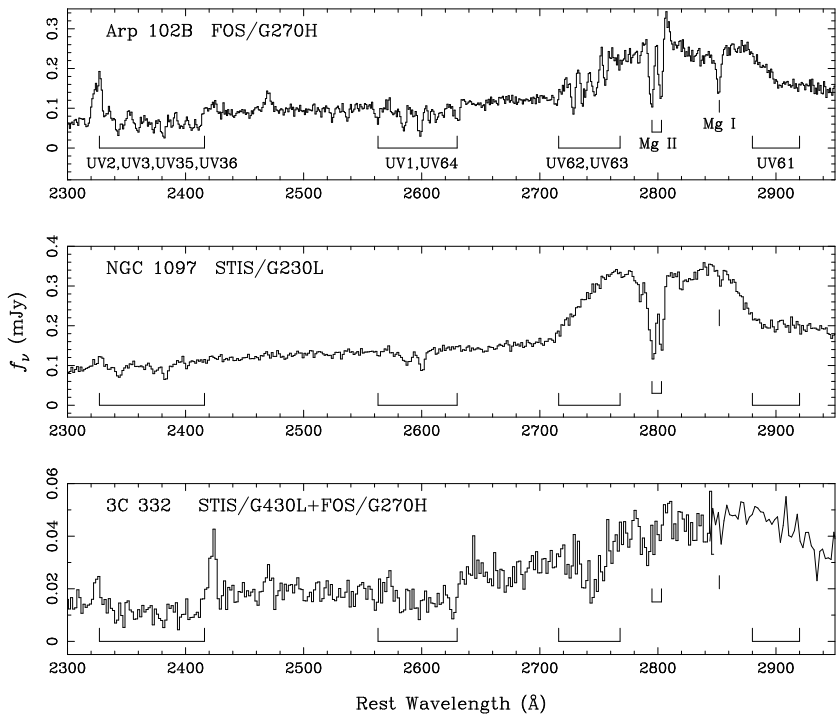}}
\caption{Comparison of the near-UV spectra of Arp~102B, 3C~332, and
NGC~1097. All three objects have \ion{Fe}{ii} and \ion{Mg}{ii}
absorption lines, although these lines seem to vary in strength from
object to object. In particular, the UV62 and UV63 multiplets are
noticeably absent in NGC~1097. These are also the objects that sport a
weak UV continuum as well as weak far-UV emission lines.}
\end{figure}

We also discovered \ion{Fe}{ii} and \ion{Mg}{ii} associated
absorption-line complexes in the UV spectra of some of the
double-peaked emitters, as shown in Figure~2.  Particularly
interesting are the \ion{Fe}{ii} UV35--36 and UV61--64 multiplets,
which arise from metastable lower levels located about 1~eV above the
ground state. The absorption lines are blueshifted by about 100--200
km s$^{-1}$ relative to the systemic velocity, which suggests that
they originate in an outflow. The case of Arp~102B was studied in
great detail by \cite[Eracleous et al. (2003)]{eracleous03} who used
X-ray observations to constrain the column density of the absorber and
constructed photoionization models to investigate its structure.  One
of the main conclusions of that work was that the absorber is very
likely to be a dense, low-ionization medium ($n\gsim 10^{11}~{\rm
cm}^{-3}$, $10^{-3.5} < U < 10^{-2.5}$), located just above the
line-emitting portion of the accretion disk. Moreover, the absorber is
likely to have the form of dense filaments, perhaps embedded in a
lower-density, higher-ionization outflow. Thus the absorption lines
trace an outflowing wind from the accretion disk, which is probably
connected to the medium responsible for the single-peaked far-UV
emission lines.

\section{Broader Implications}

The UV spectra of double-peaked emitters also lead to a number of
general insights into AGN broad-line regions and they suggest a way of
connecting double-peaked emitters with the greater AGN population. In
this scheme, which is illustrated schematically in Figure~3, luminous
AGNs (such as Seyfert galaxies and quasars) accrete at a rate that is
a substantial fraction of the Eddington rate ($\REdd\gsim 0.1$). Thus
they produce a significant wind, which is also the predominant source
of broad emission lines (\cite{murray95}; \cite{proga00}).  The
double-peaked emitters, on the other hand, are the segment of the AGN
population in which the accretion rate relative to the Eddington rate
is extremely low ($\REdd\lsim10^{-3}$). In this extreme, the inner
accretion disk turns into an ion torus, the wind diminishes, and the
accretion disk is unveiled.

We can exploit our direct view of the disk in double-peaked emitters
to study its dynamics. This can be accomplished through long-term
variability studies in which the variations of the double-peaked line
profiles are compared with models of dynamical phenomena in the disk.
Such studies require patience and persistence, as shown in the
examples presented by Lewis et al. and Gezari et al. in this volume.

\begin{figure}
\centerline{\psfig{file=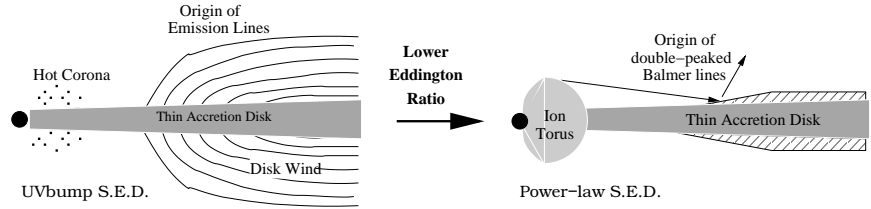,width=4.5in}}
\caption{A sketch of how the structure of the accretion disk and its
  associated wind change as the Eddington ratio goes from high values
  ($\REdd\gsim0.1$) to low values ($\REdd\lsim10^{-3}$). In the latter 
case the wind still exists but it is rather thin and inhomogeneous.}
\end{figure}

\begin{acknowledgements}
This work is supported by NASA through grant NAG5-10817 and grant
GO-08684.01-A from the Space Telescope Science Institute, which is
operated by AURA, Inc., under NASA contract NAS5-26555.
\end{acknowledgements}

\end{document}